\documentclass{article}
\usepackage{amsmath}
\usepackage{amssymb}
\usepackage{latexsym}
\begin{document}

\title{Spontaneous Symmetry Breaking and Tunneling in de Sitter Space.}
\author{Gordon L. Kane\\Michigan Center for Theoretical Physics\\
University of Michigan,\\ Ann Arbor,\\ Michigan 48109,\\ USA.\\ \\
Malcolm J. Perry\\
Department of Applied Mathematics and Theoretical Physics,\\
Centre for Mathematical Sciences,\\University of Cambridge,\\
Wilberforce Road,\\
Cambridge CB3 0WA,\\ England.\\ \\
Anna N. \.Zytkow,\\ Institute of Astronomy,\\ 
University of Cambridge,\\ Madingley Road,\\ Cambridge CB3 0HA,\\
England.}

\maketitle

\begin{abstract}
            Motivated by our earlier argument that the apparent large
cosmological constant from quantum fluctuations is actually an artifact
of not using a full quantum mechanical superposition to determine the
ground state in which the universe lives in the de Sitter space at the
beginning of inflation, we calculate the tunneling probability for the
two-well potential for a scalar field in de Sitter space.  
We include coupling of
the potential to gravity, and the effective potential from quantum
corrections.  The results show the eigenstates are the sum and
differences of the wave functions for the separate wells, i.e. a full
superposition, and the energy levels are split, with tunneling between
them determined by the Hawking-Moss instanton and not suppressed.  

\end{abstract}

\section{Introduction}

The cornerstone of our understanding of the standard model of particle physics
is the Higgs mechanism. The familiar picture in Minkowski space \cite{CW:ftn}, 
starts with a real scalar field and an associated
renormalizable potential with two degenerate minima. As is well known,
the scalar field sits at either of the two classical minima of this potential.
Its expectation value can then give a mass to gauge bosons.
The above  classical  picture can be invalidated by various quantum
effects. For example, if the temperature is sufficiently high,  \cite{DJ:pr}, 
\cite{SW:pr}, \cite{KL:a}, \cite{KL:b} or if external fields are sufficiently
strong, \cite{GS:ap}, then symmetry can be restored. Consequently, we
are motivated to explore what happens if, instead, the effect is examined
in de Sitter space. One reason is the present observed small positive 
cosmological constant which means that the universe will ultimately
be described by de Sitter 
space.  Another reason to consider this is that during
inflation, the universe is described by de Sitter space.

In an earlier work, \cite{KPZ:prep} we proposed an approach to how 
to understand the problem of large quantum fluctuations that give rise 
to an apparent cosmological constant many orders of magnitude larger than
that observed. One crucial assumption used in that approach was that
tunneling between minima of a potential in de Sitter space is not
suppressed by an effectively infinite potential barrier as it would be
in Minkowski space. In this present paper, we have established that
the simple arguments used there are borne out by a more sophisticated
treatment based on proper quantum field theoretic approach to the problem.

One might be tempted to suspect that things could 
easily be quite different from Minkowski space because of the existence of the 
Gibbons-Hawking background temperature, \cite{GH:pr}.
 The temperature of  de Sitter 
space is not a freely specifiable quantity. If the de Sitter radius
is $a$, then the Gibbons-Hawking temperature is 
\begin{equation}
T_{GH}=\frac{1}{2\pi a}
\end{equation}
So if $a$ is sufficiently small, we could expect the symmetry to be restored. 
This requires the computation of the effective potential, and we will discuss
this in detail in section four. There is also another effect that is
rather surprising. In Minkowski spacetime, the two distinct vacua are separated
by an infinite potential barrier. Although the potential energy density
separating the two minima is finite, there is no tunneling between
the two minima. This is because the volume of space of space is infinite,
and so the total energy required for a tunneling process is therefore
infinite. However, de Sitter space has finite volume, and this leads
to a non-zero tunneling amplitude. This phenomenon
is discussed and explored in section three.

\section{de Sitter Space, Temperature and Instantons.}

Our starting point is a real scalar field $\phi$ coupled to a classical 
gravitational field described by a metric $g_{ab}$. 
\footnote{Our conventions are
as follows: the metric signature is $(-+++),$ and the scalar curvature of 
de Sitter space,
(as opposed to anti-de Sitter space) is positive. The cosmological 
constant of de Sitter space is positive 
which corresponds to a positive energy density.}
The scalar field has  potential
\begin{equation}
V(\phi) = \frac{1}{2}m^2\phi^2 + \frac{1}{24}\lambda\phi^4 + 
\frac{3}{2}\frac{m^4}{\lambda}
\end{equation}
with $m^2<0$ and $\lambda > 0$. This is a double well potential with
minima at $\phi_{\pm} = \pm(6m^2/\lambda)^{1/2}$ where $V(\phi_{\pm})=0.$ 
There is also a local maximum at $\phi=0$ where $V(0)=\frac{3m^4}{2\lambda}$.
In Minkowski spacetime, the field will sit in either of the two minima.
One does not need to worry about tunneling between the two minima because
the potential barrier between the two is infinite. Whilst the energy density 
at the maximum is finite and equal to $\frac{3m^4}{2\lambda}$, the total 
energy required to move from one minimum to another
is infinite because the volume of space is infinite. A consequence of this
that we never need to worry about the possibility of the true state of the 
theory being in some superposition of the two minima.

If we couple this model to gravity, then the situation becomes 
a little  more complicated. The action for the coupled system of fields is 
\begin{equation} 
I= \frac{1}{16\pi G_N}\int g^{1/2}d^4x\ \  (R-2\Lambda)  -
\int g^{1/2}d^4x\ \  (\frac{1}{2} \nabla_a\phi\nabla^a\phi + V(\phi)). 
\end{equation}
In the above equation $G_N$ is Newton's gravitational constant, $R$ the 
Ricci scalar and $\Lambda$ the cosmological constant. One feature of this
system is that one could add a constant, $V_1$ say, to the 
potential of the scalar field without changing its equations of motion. 
Such an addition corresponds to a change in  the vacuum energy of the
scalar. In the absence of gravitation, this change has no effect.
However, in the presence of gravitation, the effect is the same as adding
$8\pi G_N V_1$ to the cosmological constant. The first to 
worry about this particular effect was Veltman \cite{MV:prl}, and it was
subsequently used to great effect by Guth \cite{AG:infl} and subsequent
investigators of inflation. We keep  $V_1=0$ but acknowledge the possible
effects of the vacuum energy of scalars through  contribution to $\Lambda.$ 
Since we are mainly 
interested in de Sitter space, we will take $\Lambda>0$.
The ambiguity between a genuine cosmological constant, and the energy density
of the vacuum, means that one can always re-interpret a cosmological constant
 $\Lambda$ as a vacuum energy density $\varrho=\frac{\Lambda}{8\pi G_N}$.

There are now three maximally symmetric solutions to the coupled
Einstein-scalar equations
derived from this action. They each have $\phi={\text const}$ in
a spacetime which is de Sitter space.
There are two stable solutions with $\phi=\phi_\pm,
\ V(\phi)=0$,
and with de Sitter radius 
\begin{equation}
a=a_\pm=\sqrt{\frac{3}{\Lambda}}.\label{eq:des1}
\end{equation}
There is also 
an unstable solution with $\phi=0,\ V(\phi)=V_0$ and 
where the de Sitter radius is 
\begin{equation}
a=a_0=
\sqrt{\frac{3}{\Lambda+{8\pi G_N V_0}}} \label{eq:des2}
\end{equation}
with $V_0=\frac{3m^4}{2\lambda}$.
The metric of the
spacetime is given by
\begin{equation}
ds^2 = -dt^2 + a^2\cosh^2\biggl({\frac{t}{a}}\biggr)d\Omega_3^2
\end{equation}
where $a$ is $a_\pm$ or $a_0$ as appropriate.
$d\Omega_3^2$ is the  metric on a symmetric unit $S^3$ of volume
$2\pi^2$. This is the metric on de Sitter space \cite{WdeS:cosm}.
At any instant of cosmological time $t$, space is described by 
an $S^3$ of radius $a\cosh(t/a)$.
Note however that the volume of space is always  finite, and  arguments 
against the relevance of tunneling based on flat space reasoning
no longer apply. 

Firstly, we review and 
examine the situation in which there is just the gravitational
field. We quantize this system by using Euclidean path integral techniques.
The Euclidean action is
\begin{equation}
I_E = \frac{-1}{16\pi G_N}\int g^{1/2}d^4x (R-2\Lambda) 
\end{equation}
In Euclidean gravity, one considers the partition function
\begin{equation}
Z=\int {\cal D}[g] e^{-I_E[g]}
\end{equation}
where the path integral is taken over all metrics $g$ (possibly subject to 
some boundary conditions). We can evaluate $Z$ in the loop expansion. 
The lowest order contribution to $Z$ is just the tree-level one where one
evaluates $Z$ by substituting  the classical Euclidean spacetime metric
into the action. The classical spacetimes obey the Euclidean Einstein 
equations
\begin{equation}
R_{ab}=\Lambda g_{ab}.
\end{equation}
The solution of lowest action, and hence the most important contribution
to $Z$ has $\phi=0$ and metric
\begin{equation}
ds^2 = (1-\frac{1}{3}\Lambda r^2)d\tau^2 + \frac{dr^2}{1-\frac{1}{3}\Lambda
r^2}+r^2(d\theta^2+\sin^2\theta d\phi^2)
\end{equation}
This is the metric on a symmetric $S^4$. It should be noted that this
metric requires one to make $\tau$ a periodic coordinate with period
$2\pi\sqrt{\frac{3}{\Lambda}}=2\pi a$.
Evaluating the action for this
solution gives
\begin{equation}
I_E=-\frac{3\pi}{\Lambda}=-\frac{\pi a^2}{G_N}
\end{equation}
As has been emphasized many times, the periodicity of the Euclidean time
coordinate $\tau$ leads to the background Gibbons-Hawking temperature given by
\begin{equation}
T_{GH}=\frac{1}{2\pi a}.
\end{equation}

The physical interpretation of this result is easy to understand. 
Geodesic observers in de Sitter space can easily describe their observations
using static coordinates. Static coordinates in de Sitter space give a
metric
\begin{equation}
ds^2 = -(1-\frac{1}{3}\Lambda r^2)dt^2 + \frac{dr^2}{(1-\frac{1}{3}
\Lambda r^2)}+r^2d\theta^2 + r^2\sin^2\theta d\phi^2
\end{equation}  
where $(r,\theta,\phi)$ are polar coordinates and $t$ is a time coordinate.
These coordinates however do not cover the entirety of the spacetime. 
Were one to analytically continue this spacetime to a Euclidean signatured
space by setting $t=-i\tau,$ one would arrive at the metric on 
$S^4$ given above.
A geodesic observer moves along the line $r=0$ where $t$ coincides
with the observers proper time. All such inertial observers are equivalent
since de Sitter spacetime is maximally symmetric. Under the action 
of the de Sitter group, one observer's world-line can be transformed into
any other one. In every case, it is possible to construct a set of static 
coordinates
around that observers world line. In those coordinates, the metric is
always given by that above.

The surface $r=\sqrt{\frac{3}{\Lambda}}=a$ as $t\rightarrow\infty$
is a null surface and represents the cosmological horizon of 
the observer. Note however that different observers will have different 
horizons. The Gibbons-Hawking temperature results from the Hawking
temperature of the cosmological horizon. With this horizon, there is
an associated entropy, $S$. It can be found by looking at the
the Hamiltonian, $H$ for this particular observer. Suppose that
the inverse temperature $\beta = T^{-1}.$ The canonical 
partition function for the gravitational field for the this observer is
\begin{equation}
Z= tr\ e^{-\beta H}
\end{equation}
The partition function is precisely the object found from the
gravitational path integral.
The entropy is $S$ can then be determined from
\begin{equation}
S=-\frac{\partial}{\partial T}(T \ln Z)
\end{equation}
Evaluating $S$ gives
\begin{equation}
S=\frac{\pi a^2}{G_N}
\end{equation}
which agrees with the general Hawking expression for the entropy of
\begin{equation}
S = \frac{A}{4G_N}
\end{equation}
where $A$ is the area of the event horizon. The idea that the Hamiltonian 
defined by the Euclidean continuation is the correct one for inertial
observers is confirmed by the consistency of this picture. 

The same techniques of Euclidean field theory can be applied
with minor modification to the
scalar field theory coupled to gravity that 
we are interested in here.
The Euclidean action is now
\begin{equation}
I_E=\frac{-1}{16\pi G_N}\int g^{1/2}d^4x (R-2\Lambda) + \int g^{1/2}d^4x
(\frac{1}{2}\partial_a\phi\partial^a\phi+V(\phi))
\end{equation}
This yields the field equations
\begin{equation}
R_{ab}-\frac{1}{2}Rg_{ab}+\Lambda g_{ab}=8\pi G_N(\partial_a\phi\partial_b\phi
-\frac{1}{2}g_{ab}\partial_c\phi\partial^c\phi + g_{ab}V(\phi))
\end{equation}
Our aim is to determine the true vacuum state of the theory, and 
determine its energy density. The easiest way to measure the energy density
in de Sitter spacetime is just to measure the de Sitter radius, $a_E$.
One can then infer the effective cosmological constant $\Lambda_E$ since
\begin{equation}
\Lambda_E=\frac{3}{a_E^2}.
\end{equation}
Suppose we are in a situation in which the true cosmological constant
is $\Lambda$ but the observed cosmological constant was $\Lambda_E$.
Then the vacuum energy density $\varrho$ that we could attribute to 
the tunneling process is just
\begin{equation}
\varrho=\frac{\Lambda_E-\Lambda}{8\pi G_N}.
\end{equation}
These relations are just a reinterpretation of equations (\ref{eq:des1}) and
(\ref{eq:des2}) where now we use the radius to determine the energy density, 
rather than using the energy density to determine the radius. The point is 
that one can use an unambiguous geometric quantity to define what is meant by
energy density. Then, the energy density of the vacuum is determined by
the difference between the true cosmological constant $\Lambda$ and the 
effective cosmological constant $\Lambda_E$ as determined by the 
de Sitter radius $a_E$.   
In Euclidean field theory, the easiest way to measure the effective
cosmological constant is to start from the volume ${\cal V}$, \cite{SWH:vce}.
 The
volume ${\cal V}$ of Euclidean de Sitter space, $S^4$ in terms of the de Sitter
radius $a_E$ is
\begin{equation}
{\cal V}= \frac{8\pi^2}{3}a_E^4.
\end{equation}
A knowledge of ${\cal V}$ therefore allows us to calculate the energy
density.

${\cal V}$ can be found by modifying the path integral. Define
\begin{equation}
Z[\alpha]= \int {\cal D}[g] e^{(-I_E[g]-\alpha{\cal V})}.
\end{equation}
Since
\begin{equation}
{\cal V}=\int g^{1/2}d^4x
\end{equation}
the new path integral is the same as the old one but with the cosmological 
constant $\Lambda$ replaced by $\Lambda+8\pi G_N \alpha$. To evaluate
the  matrix elements of ${\cal V}$, we use the fact that
\begin{equation}
{\cal V}=-Z^{-1}\frac{\partial Z}{\partial \alpha}\Bigg\vert_{\alpha=0}
\end{equation}

There are two degenerate vacuum states in our model, corresponding to the 
two  minima
at $\phi_{\pm},$ which we call $\vert +\rangle$ and $\vert -\rangle$
respectively. We can evaluate the new path integral with these as
initial or final states, thereby giving us the matrix elements of
the operator $Z$.
The Euclidean path integral has three saddle points. The first has the scalar
fields given by $\phi_+$ and the gravitational field by Euclidean de Sitter 
space of radius $a_\pm$.
The second is similar but with the scalar being given by $\phi_-$. 
The last is with the scalar vanishing, and  the de Sitter radius now being 
given by $a_0$. If one is interested in the matrix elements
$\langle - \vert Z \vert - \rangle$ or 
$\langle + \vert Z \vert + \rangle$ then the scalar field just sits in
the classical minimum. Thus
\begin{equation}
\langle - \vert Z \vert - \rangle = \langle + \vert
Z \vert + \rangle = e^{\pi \tilde a_\pm^2}
\end{equation}
where now 
\begin{equation}
\tilde a_{\pm}^2 = \frac{3}{\Lambda+8\pi G_N\alpha} \label{eq:apm}
\end{equation}
If we try to compute $\langle \phi_- \vert Z \vert \phi_+ \rangle$
then the scalar field must start at $\phi_+$ and end at $\phi_-$.
There is no classical solution to the coupled Einstein-scalar field equations
that obeys these boundary conditions, nevertheless there 
is a saddle point of the action on the path from $\phi_-$ to $\phi_+$.
If one integrates over all $\phi$ that go from $\phi_-$ to $\phi_+$ then 
one must pass through $\phi=0$, the saddle point where the de Sitter 
radius is $a_0$. Thus, there is a non-vanishing contribution to this 
matrix element. This solution is the so-called 
\lq\lq Hawking-Moss\rq\rq instanton, \cite{HM:inst}.
Consequently, in the zero-loop
approximation we find that 
\begin{equation}
\langle + \vert  Z \vert - \rangle = \langle - \vert
Z \vert + \rangle = e^{\pi \tilde a_0^2}
\end{equation}
where now
\begin{equation}
\tilde a_0^2 = \frac{3}{\Lambda+8\pi G_N V_0 +8\pi G_N\alpha}
\label{eq:a0}
\end{equation}
Note that when $\alpha=0$, $\tilde a_{\pm}$ and $\tilde a_0$ coincide with
$a_\pm$ and $a_0$ respectively.
We see from this that the matrix elements of $Z$ are not
diagonal. Thus, matrix elements of ${\cal V}$ are not diagonal  
either.

Evaluation of $Z$ leads to 
\begin{equation}
\begin{pmatrix} e^A & e^B \\ e^B & e^A
\end{pmatrix}
\end{equation}
where $A=\pi\tilde a_\pm^2$ and $B=\pi\tilde a_0^2$. The matrix elements of
${\cal V}$ are then  
\begin{equation}
{\cal V} = \frac{1}{e^{2B}-e^{2A}}\begin{pmatrix} 
e^{2A}A_\alpha-e^{2B}B_\alpha&e^{A+B}(B_\alpha - A_\alpha) \\ 
e^{A+B}(B_\alpha-A_\alpha)&e^{2A}A_\alpha-e^{2B}B_\alpha\end{pmatrix}
\Bigg\vert_{\alpha=0}
\end{equation}
where $A_\alpha$ and $B_\alpha$ are ${\frac{\partial A}{\partial\alpha}}$
and $\frac{\partial B}{\partial \alpha}$ respectively.
The eigenvalues of ${\cal V}$ are
\begin{equation}
-\frac{(A_\alpha e^A+B_\alpha e^B)}{e^A+e^B} \quad\text{and}\quad
\frac{B_\alpha e^B - A_\alpha e^A}{e^A-e^B}
\end{equation}
The eigenvectors are $\frac{1}{\sqrt{2}}(\vert + \rangle \pm \vert - \rangle)$.
Thus the energy of the true ground state of the theory is shifted relative to 
the classical energy. Furthermore, the degeneracy of the ground state 
is lifted. This comes about because the tunneling probability between 
the two distinct classical minima does not vanish as it does in flat
 Minkowski space. This is a direct consequence of the fact that the 
volume of space is finite. Thus, we find for this problem, 
a field theoretic result which looks more like a quantum mechanical 
result than
a flat Minkowski space field theory result. This seems to
fit in well with many recent discussions of physics in de Sitter
space in which it argues that the space of states must indeed be finite,
\cite{EW:des}.

If $V_0\ll\frac{\Lambda}{8\pi G_N}$, then the energy density of these
states is $\Lambda/8\pi G_N \mp\frac{1}{2}V_0$. 
The spacing between the states grows
as one increases $V_0$ until $V_0=\Lambda/8\pi G_N$ when our
method breaks down. At this point, the lower energy state 
would have negative energy density apparently 
leading to an anti-de Sitter space. 
However,  our methods cannot be easily applied to such a situation
so we will not discuss this possibility further.

One interesting feature of this result highlights a  well-known 
objection to having the universe in a mixed state. 
If this happens, the S-matrix then fails to obey cluster decomposition.  
Should we be alarmed by this piece of folklore? 
Almost certainly not. In a closed universe, there is no notion of 
an asymptotic state. There is no S-matrix therefore. Furthermore, the failure 
of cluster decomposition because of infra-red problems is not in itself
an unphysical feature of a theory. An obvious counterexample to
the idea that cluster decomposition is an essential prerequisite of
a sensible field theory is provided by QCD. The asymptotic states are not
the elementary states of the theory. We should probably not be surprised
that in a theory of gravitation where there are long-  (or possibly infinite-)
range forces, that cluster decomposition should not necessarily apply.

\section{Effective Potential}

So far, our results are semi-classical. It could be that quantum corrections
change this picture. When quantum corrections are taken into account, the 
potential $V(\phi)$ is replaced by the effective potential $V_{eff}(\phi)$.
The vacuum states of the theory are then described by $V^\prime_{eff}(\phi)=0$
and $V^{\prime\prime}_{eff}(\phi)>0$. To calculate $V_{eff}$, start by
decomposing $\phi$ into its classical part $\phi_0$ and its quantum
part $\hat\phi$. Expanding the Lagrangian for $\phi$,  the quadratic term
is
\begin{equation}
-\hat\phi\Box\hat\phi - \frac{1}{2}\hat\phi^2V^{\prime\prime}(\phi_0)
=\hat\phi L \hat\phi
\end{equation}
which defines a second order differential operator $L$. To lowest order
(one-loop), the quantum corrections give an effective potential
\begin{equation}
V_{eff}(\phi)= V(\phi) - \frac{1}{2}{\text{ln det}}L
\end{equation}
A standard calculation gives the zero temperature flat space result.
Evaluating the effective potential in the modified minimal subtraction scheme, 
we find
\begin{equation}
V_{eff}= \frac{1}{2}m^2\phi^2 + \frac{1}{24}\lambda\phi^4
+\frac{1}{64\pi^2}\biggl(m^2+\frac{1}{2}\lambda\phi^2\biggr)^2
\biggl[\ln\biggl(\frac{m^2+\frac{1}{2}\lambda\phi^2}{\mu^2}\biggr)
-\frac{3}{2}\biggr]
\end{equation}
where $m$ and $\lambda$ are now the renormalized values of the original
mass parameter and coupling constant respectively and $\mu$ is the 
renormalization scale. Thus the first two terms on the right are the 
classical ones and the third the quantum correction.  
We must however be rather careful in interpreting
this result. Suppose that  $m^2<0$ as we originally assumed.
Then if $\phi^2<-\frac{2m^2}{\lambda},$ 
the argument of the logarithm in the quantum correction
term to the effective potential  is negative, and the effective potential
becomes complex. Thus for small fields strengths the perturbation 
expansion appears not to  work. 
The reason for this apparent difficulty is that we are in fact expanding
in powers of $\lambda\ln(m^2+\frac{1}{2}\lambda\phi^2)$ and not
as one might have thought in powers of $\lambda$.
We can hardly expect an expansion in terms of an ill-defined parameter
to give reasonable results. We will assume that quantum corrections
for small values of the fields do not make a significant difference
to the form of the effective potential. For $\phi^2>-\frac{2m^2}{\lambda}$
the expansion is well-defined and we see that these corrections 
have the effect of slightly lowering the minima of the potential
and making the sides of the well
somewhat steeper. As long as $\phi^2$ is not too large, the shape of the
potential is not dramatically changed. 
So at zero temperature, there is believed
to be no dramatic change to the form of the potential. 

Things become completely different at non-zero temperature. One can calculate
the effective potential in a high temperature expansion. The effective
potential  then gets modified by the addition of
\begin{equation}
-\frac{\pi^2}{90\beta^4}+\frac{m^2+\frac{1}{2}\lambda\phi^2}{24\beta^2}
-\frac{(m^2+\frac{1}{2}\lambda\phi^2)^{3/2}}{12\pi\beta}-
\frac{(m^2+\frac{1}{2}\lambda\phi^2)^2}{64\pi^2}
\ln(m^2+\frac{1}{2}\lambda\phi^2)\beta^2+\ldots
\end{equation}
where $\beta$ is inverse temperature, $\beta=1/T$.
Thus, as was shown in \cite{KL:a},\cite{KL:b}, \cite{SW:pr}, and \cite{DJ:pr}
for $T>T_c=\sqrt{\frac{-24m^2}{\lambda}}$ the naive vacuum with $\phi=0$
is the correct one. There is a phase transition in the theory.
Above $T_c$, the effective potential has only a single minimum
at $\phi=0$, and below $T_c$ there are minima with $\phi\ne 0$.

So, if the temperature is sufficiently high, the possibility of having
the ground state of the theory being a mixed state is wiped out by
large thermal fluctuations. de Sitter space has its own temperature 
because of its horizons so that we need to investigate
the effective potential 
in de Sitter space to see if the effect we are interested in survives.
The effective potential for de Sitter space has been calculated in
\cite{GS:ap}, and  \cite{BA:np}. 

The principal difference between the flat space calculation and the de
Sitter space one lies in the nature of the differential operator $L$.
In flat space, $L$ has a continuous spectrum and familiar  Fourier
transform techniques allow its computation to be straightforward. 
In euclidean de Sitter space $L$ has a discrete spectrum. Putting
$M^2=m^2+\frac{1}{2}\lambda\phi^2$, we see that the eigenvalues of
$L$ are $\lambda_n$ with degeneracy $d_n$,
\begin{equation}
\lambda_n=\frac{1}{a^2}n(n+3)+M^2,\ \ \ \ d_n=\frac{1}{6}(n+1)(n+2)(2n+3)
\end{equation}
where $a$ is the de Sitter radius.
If we define the zeta function of $L$ to be
\begin{equation}
\zeta(s,L)=\sum_{n=0}^{\infty}d_n\lambda_n{}^{-s}
\end{equation} 
then the zeta function is convergent for $Re(s)>2$. It has a unique analytic 
continuation to the entire complex plane except for simple poles at $s=1$ and
$s=2$. In terms of the zeta function, the quantum correction to the potential
is
\begin{equation}
-\frac{3}{8\pi^2a^4}\biggl(-\frac{1}{2}\zeta^\prime(0,L)+\zeta(0,L)\ln\mu
\biggr).
\end{equation}
$\zeta(0,L)$ can be evaluated exactly and is
\begin{equation}
\zeta(0,L)=\frac{1}{3}\zeta_H(-3,\frac{3}{2})-
\frac{1}{24}\zeta_H(-1,\frac{3}{2})+\frac{1}{24}X+\frac{1}{12}X^2
\end{equation}
where $\zeta_H$ is the Hurwitz zeta function
\begin{equation}
\zeta_H(s,\alpha)=\sum_{n=0}^\infty (n+\alpha)^{-s}
\end{equation}
and
\begin{equation}
X=M^2a^2-\frac{9}{4}.
\end{equation}
Explicit evaluation gives $\zeta_H(-3,\frac{3}{2})=-\frac{127}{960}$
and $\zeta_H(-3,\frac{1}{2})=-\frac{11}{24}$.
Similarly, one can calculate $\zeta^\prime(0,L)$
and it is
\begin{equation}\begin{split}
\zeta^\prime(0,L)&=\frac{2}{3}\zeta_H^\prime(-3,\frac{3}{2})
-\frac{1}{6}\zeta_H^\prime(-3,\frac{1}{2})-\frac{1}{72}X+\frac{1}{12}X^2\\
&+\psi(\frac{1}{2})(-\frac{1}{12}X-\frac{1}{6}X^2)\\
&+\frac{1}{24}\int_0^\infty\frac{dt}{t^4} \text{cosech}(\frac{1}{2}t)
\biggl(\sqrt{X}\sin(\sqrt{X}t)(8Xt^3-46t)\\&+\cos(\sqrt{X}t(-46-24Xt^2)+
(46-Xt^2-2X^2t^4)\biggr)\\
\end{split} \label{veffindes}
\end{equation}
where $\psi(z)$ denotes the logarithmic derivative of the gamma function.
This form of the effective potential closely
resembles a  Schwinger-type proper time
formula.
As one would expect, the singularity in the integral at $t=0$ is removable.
As $a\rightarrow\infty$ this expression coincides with the
flat space result. Thus, if the de Sitter radius is sufficiently large and
also provided that the  hump in the potential is sufficiently small, 
the effective potential will be unchanged
compared to the flat space case. The hump must remain small compared to
the de Sitter radius  so that there can 
be no significant backreaction of the potential hump onto
the effective cosmological constant thereby altering the de Sitter
radius significantly between the minimum of the potential and 
the top of the hump. Therefore, under these circumstances 
the true ground state of the
theory will be a mixed state. In the  more general case, it is rather 
complicated to evaluate in detail the effective potential and draw any
definite conclusions. However, Allen \cite{BA:np} did study the effective 
potential numerically and concluded that if $Ma\lesssim 8$, 
then the effective
potential had only a single minimum at the origin. Lastly, it is simple
to evaluate the effective potential in the limit as $a\rightarrow 0$
where we find
\begin{equation}
V_{eff}(\phi)= \frac{1}{2}m^2\phi^2+\frac{1}{24}\lambda\phi^4
+A(\mu)(m^2+\frac{1}{2}\lambda\phi^2)a^{-2}
\end{equation}
This can be deduced from (\ref{veffindes}) 
by taking the limit as $a\rightarrow 0$, The first term in such an expansion 
is just proportional to $a^{-4}$, the cosmological constant, and so does not
contribute to the effective potential. The next term is proportional to
$M^2a^{-2}$ but with a $\mu$ dependent coefficient. Thus some renormalization
prescription will be necessary to determine the coefficient
However, assuming $A(\mu)>0$, so that the theory is stable, 
this term will always
overwhelm the other terms and so there will be a single minimum.
Thus for sufficiently high temperature, there will always be a
single minimum. We therefore see that our intuition of what the
effective potential must be, based on flat space
reasoning, turns out to be correct. Nevertheless, because of the complexity
of (\ref{veffindes}),
it seems to us that further investigations of the effective potential could
well be quite enlightening.

\section{Unequal Minima and Tunneling}

The situation we have discussed has the minima of its potential precisely
degenerate. Such a situation is unlikely to be realized in practice. So
what happens if the minima are no longer precisely degenerate? We will not 
attempt to present a complete discussion of this point. However, suppose
that the de Sitter radii are now $a_+$ and $a_-$ for the minima and
$a_0$ for the maximum, then one replaces the matrix $Z$ by
\begin{equation}
\begin{pmatrix} e^A & e^B \\ e^B & e^C
\end{pmatrix}
\end{equation}
where now $A=\pi a_+^2, B=\pi a_0^2$ and $C=\pi a_-^2$. The corresponding 
elements of ${\cal V}$ are
\begin{equation}
{\cal V} = \frac{1}{e^{2B}-e^{A+C}}\begin{pmatrix} 
e^{A+C}A_\alpha-e^{2B}B_\alpha&e^{B+C}(B_\alpha - C_\alpha) \\ 
e^{A+B}(B_\alpha-A_\alpha)&e^{A+C}C_\alpha-e^{2B}B_\alpha\end{pmatrix}
\Bigg\vert_{\alpha=0}
\end{equation}
where $A_\alpha$ and $B_\alpha$ are ${\frac{\partial A}{\partial\alpha}}$
and $\frac{\partial B}{\partial \alpha}$ respectively.
Then eigenvalues of ${\cal V}$ are
\begin{equation}
\frac{(A_\alpha e^{A+C}-B_\alpha e^{2B})}{e^{2B}-e^{A+C}} \quad\text{and}\quad
\frac{(C_\alpha e^{A+C}-B_\alpha e^{2B})}{e^{2B}-e^{A+C}}
\end{equation}
It is easy to see qualitatively what happens. As the two minima separate,
the eigenvalues becomes closer and closer to $8\pi G_N a_{\pm}^4/3$. 
These are what one would expect as the tunneling between the two minima
is switched off. The corresponding eigenvectors are then more and 
more concentrated about the two separate minima. 
Thus, as the separation becomes larger, the true ground state becomes closer
and closer to the bottom of the lowest potential minimum. 
However, as long as $A-C \ll A,B,C,$ the results are not qualitatively 
different from
the degenerate case, so the application of our results remains valid for a
potential landscape in which the potential depths vary somewhat.
However, potentials  with very different minima would not contribute
significantly to any superposition.

This raises
the important cosmological question of how rapid the transitions are between
the two different states. Such questions have had their answers  sketched,
mainly in the context of inflationary scenarios,
in the
works of Callan and Coleman \cite{CC:pr}, Coleman \cite{C:pr}, 
Coleman and DeLuccia \cite{Cd:pr}, Hawking and Moss \cite{HM:inst},
Steinhardt \cite{PS:infl} and Linde \cite{AL:infl} amongst others.
We will return in a separate publication to this
particular issue.

\section{Acknowledgments}

We would like to thank Finn Larsen and Mike Douglas
for interesting conversations
during the course of this work. Two of us (MJP, ANZ) would like
thank MCTP for its generous support and hospitality during the course
of this work.
 This research was supported in part
by the US Department of energy.


\begin{thebibliography}{99}
\bibitem{CW:ftn} S. Coleman and E. Weinberg,
 ``Radiative Corrections as the Origin of Spontaneous
Symmetry Breaking,'' Phys. Rev D7: 1888-1910,1973.

\bibitem{DJ:pr} L. Dolan and R. Jackiw, ``Symmetry Behavior at Finite
Temperature,'' Phys. Rev. D9: 3320-3341, 1974.

\bibitem{SW:pr} S. Weinberg, ``Gauge and Global Symmetries at High 
Temperature,'' Phys. Rev.D9: 3357-3378, 1974.

\bibitem{KL:a} D.A. Kirzhnits and A.D. Linde, 
``A Relativistic Phase Transition,'' Sov. Phys. JETP, 40, 628, 1975.

\bibitem{KL:b} D.A. Kirzhnits and A.D. Linde,
Symmetry Behavior in Gauge Theories, Ann. Phys. (NY), 101, 195-238, 1976.

\bibitem{GS:ap} G. M. Shore, ``On the Meissner Effect in Gauge Theories,''
Ann Phys. (NY), 134, 259, 1981.

\bibitem{KPZ:prep} G.L. Kane, M.J. Perry and A.N. \.Zytkow, ``A Possible
Mechanism for Generating s Small Positive Cosmological Constant,'' 
hep-th 0311152.

\bibitem{MV:prl} M. Veltman, ``Cosmology and the Higgs Mass,''
Phys. Rev. Lett. 34, 777, 1975.

\bibitem{AG:infl} A.H. Guth, ``The Inflationary Universe: A Possible
Solution to the Horizon and Flatness Problems,'' Phys. Rev. D23, 347-356, 1981.

\bibitem{WdeS:cosm} W. de Sitter, ``On the Curvature of Space,''
Proc. Kon. Ned. Akad. Wet. 20, 229-243, 1917.

\bibitem{GH:pr} G.W. Gibbons and S.W. Hawking, ``Cosmological Event Horizons, 
Thermodynamics and Particle Creation,'' Phys. Rev. D15, 2738-2751, 1977.

\bibitem{SWH:vce} S.W. Hawking, ``Space-Time Foam,'' Nucl. Phys. B144, 
349-362, 1978.

\bibitem{HM:inst} S.W. Hawking and I.G. Moss, ``Supercooled Phase Transitions
in the Very Early Universe,'' Phys. Lett. B110, 35, 1982.

\bibitem{EW:des} E. Witten, ``Quantum Gravity in de Sitter Space,''
hep-th 0106109.

\bibitem{GS:apa} G.M. Shore, ``Radiatively Induced Spontaneous Symmetry 
Breaking and Phase Transitions in Curved Space-time,'' Ann. Phys.  (NY),
128, 376, 1908.

\bibitem{BA:np} B. Allen, ``Phase Transitions in de Sitter Space,''
Nucl. Phys. B226, 228-252, 1983.

\bibitem{C:pr} S. Coleman, ``The Fate of the False Vacuum: 1 - 
Semiclassical Theory,'' Phys. Rev. D15, 2929-2936, 1977.

\bibitem{CC:pr} C. Callan and S. Coleman,
``The Fate of the False Vacuum: 2. First Quantum Corrections,'' 
Phys. Rev. D16, 1762-1768, 1977.

\bibitem{Cd:pr} S. Coleman and F. deLuccia,  `` Gravitational Effects
on and of Vacuum Decay,'' Phys. Rev. D21, 3305, 1980.

\bibitem{PS:infl} A. Albrecht and P. Steinhardt, ``Cosmology for Grand Unified
Theories with Radiatively Induced Symmetry Breaking,'' Phys. Rev. Lett. 48,
1220-1223, 1982.

\bibitem{AL:infl} A.D. Linde, ``A New Inflationary Universe Scenario: 
A Possible Solution of the Horizon, Flatness, Homogeneity, Isotropy and
Primordial Monopole Problems,'' Phys. Lett. B108, 389-393, 1982. 

\end{thebibliography}
\end{document}